\documentclass[twocolumn]{aastex631mod}
\usepackage{threeparttable}
\usepackage{multirow}
\usepackage{comment}
\usepackage{url} 
\usepackage{hyperref}
\usepackage{enumitem}
\usepackage{multirow}
\usepackage{booktabs}
\usepackage{amsmath,amssymb}
\usepackage{savesym}
\savesymbol{tablenum}
\usepackage{siunitx}
\restoresymbol{SIX}{tablenum}
\usepackage{hyperref}
\usepackage[nameinlink,noabbrev]{cleveref}
\usepackage[normalem]{ulem}
\usepackage{graphicx}

\makeatletter
\usepackage{etoolbox}
\patchcmd\H@refstepcounter{\protected@edef}{\protected@xdef}{}{}
\makeatother

\definecolor{Green}{HTML}{00A64F}

\shorttitle{APOGEE DR17 RV standard stars}
\shortauthors{Li et al.}

\begin{document}

\title{An update of the catalog of radial velocity standard stars from the APOGEE DR17}

\correspondingauthor{Yang Huang}
\email{huangyang@ucas.ac.cn}

\author[0000-0002-4033-2208]{Qing-Zheng Li}
\affil{Yunnan Observatories, Chinese Academy of Sciences, Kunming, Yunnan 650011,  People's Republic of China}
\affil{School of Astronomy and Space Science, University of Chinese Academy of Sciences, Beijing 100049,  People's Republic of China}

\author[0000-0003-3250-2876]{Yang Huang}
\affil{School of Astronomy and Space Science, University of Chinese Academy of Sciences, Beijing 100049,  People's Republic of China}
\affil{Key Lab of Optical Astronomy, National Astronomical Observatories, Chinese Academy of Sciences, Beijing 100101,  People's Republic of China}

\author[0000-0002-2449-9550]{Xiao-Bo Dong}
\affil{Yunnan Observatories, Chinese Academy of Sciences, Kunming, Yunnan 650011,  People's Republic of China}



\begin{abstract}

We present an updated catalog of 46,753 radial velocity (RV) standard stars selected from the APOGEE DR17. These stars cover the Northern and Southern Hemispheres almost evenly, with 62\% being red giants and 38\% being main-sequence stars. These RV standard stars are stable on a baseline longer than 200 days (54\% longer than one year and 10\% longer than five years) with a median stability better than 215\,m\,s$^{-1}$. The average observation number of those stars are 5 and each observation is required to have spectral-to-noise-ratio (SNR) greater than 50 and RV measurement error smaller than 500 m\,s$^{-1}$. Based on the new APOGEE RV standard star catalog, we have checked the RV zero points (RVZPs) for current large-scale stellar spectroscopic surveys including RAVE, LAMOST, GALAH and {\it Gaia}. By carefully analysis, we estimate their mean RVZP to be $+0.149$\,km\,s$^{-1}$, $+4.574$\,km\,s$^{-1}$ (for LRS), $-0.031$\,km\,s$^{-1}$ and $+0.014$\,km\,s$^{-1}$, respectively, for the four surveys. In the RAVE, LAMOST (for MRS), GALAH and {\it Gaia} surveys, RVZP exhibits systematic trend with stellar parameters (mainly [Fe/H], $T_{\rm{eff}}$, log\,$g$, $G_{\rm{BP}}-G_{\rm{RP}}$ and $G_{\rm{RVS}}$). The corrections of those small but clear RVZPs are of vital importances for these massive spectroscopic surveys in various studies that require extremely high radial velocity accuracies.

\end{abstract}



\section{Introduction} \label{section:intro}

The velocity component of a star in the line-of-sight direction can be defined by the {\it Doppler shift} of the spectrum captured by the telescope. It can be converted into the framework of the Solar System's center of mass, called the ``barycentric" or ``heliocentric" radial velocity (RV). The ``barycentric" or ``heliocentric" RV represents the rate of change of the distance between the Sun and the star \citep[for the detail definition of RV, see][]{2003A&A...401.1185L}. The measurement of RV is essential to the construction of a complete stellar 6D information (3D position and 3D velocity). Its accuracy is required to be better than several km\,s$^{-1}$, or even a few m\,s$^{-1}$, for various Galactic studies such as understanding the structure and assembly history of the Milky Way \citep{1987gady.book.....B,2012ApJ...753..148B,2015ApJS..216...29B,2018MNRAS.478..611B,2018A&A...616A..11G,2018Natur.563...85H,2020ARA&A..58..205H}, estimating the mass of the Milky Way \citep{2008ApJ...684.1143X,2012ApJ...759..131B,2016MNRAS.463.2623H,2019ApJ...871..120E,2022arXiv221210393Z}, defining orbital parameters and characteristics of binary systems \citep{1995Natur.378..355M,2017MNRAS.469L..68G,2018MNRAS.476..528E,2018RAA....18...52T,2019Natur.575..618L,2022ApJ...933..119L}, identification of exoplanets \citep{2016PNAS..11311431X,2020A&A...636A..74T} and systematic searching for hypervelocity stars \citep{2005ApJ...622L..33B,2014ApJ...787...89B,2017ApJ...847L...9H,2021ApJ...907L..42H,2020MNRAS.491.2465K,2021ApJS..252....3L,2022ApJ...933L..13L,2022MNRAS.515..767M,2022arXiv220704406L}.

In the past decades, RVs have been measured for over tens of million stars from a series of large-scale spectroscopic surveys, including, the ground-based surveys, such as the GALAH \citep{2015MNRAS.449.2604D,2018MNRAS.478.4513B}, the SDSS/APOGEE \citep{2017AJ....154...94M,2017AJ....154...28B,2022ApJS..259...35A}, the RAVE \citep{2006AJ....132.1645S,2017AJ....153...75K}, the SDSS/SEGUE \citep{2009AJ....137.4377Y,2011AJ....142...72E,2013AJ....145...10D,2022ApJS..259...60R}, the LAMOST \citep{2012RAA....12..723Z,2012RAA....12..735D,2012RAA....12.1197C,2015RAA....15.1095L}, and the space-based surveys, i.e., the Gaia-RVS \citep{2004MNRAS.354.1223K,2018A&A...616A...5C,2018A&A...616A...1G,2018A&A...616A..11G}. In the near future, more stellar RVs will be obtained thanks to the ongoing/planing massive spectroscopic surveys, such as the SDSS-V \citep{2017arXiv171103234K}, 4MOST \citep{2019Msngr.175....3D}, and DESI \citep{2016arXiv161100036D,2019AJ....157..168D}.

The measurement of RV can be influenced by various factors, including the type of instrument, the spectral resolution, the accuracy of wavelength calibration, the methodology used to derive RV, and even observation conditions and environments. These factors can lead to significant variations in RV measurements. To correct for these effects, it is necessary to construct a set of RV standard stars which are stable enough in a long observation baseline.

\defcitealias{2018AJ....156...90H}{H18}

\begin{figure} 
\centering
\includegraphics[scale=0.09,angle=0]{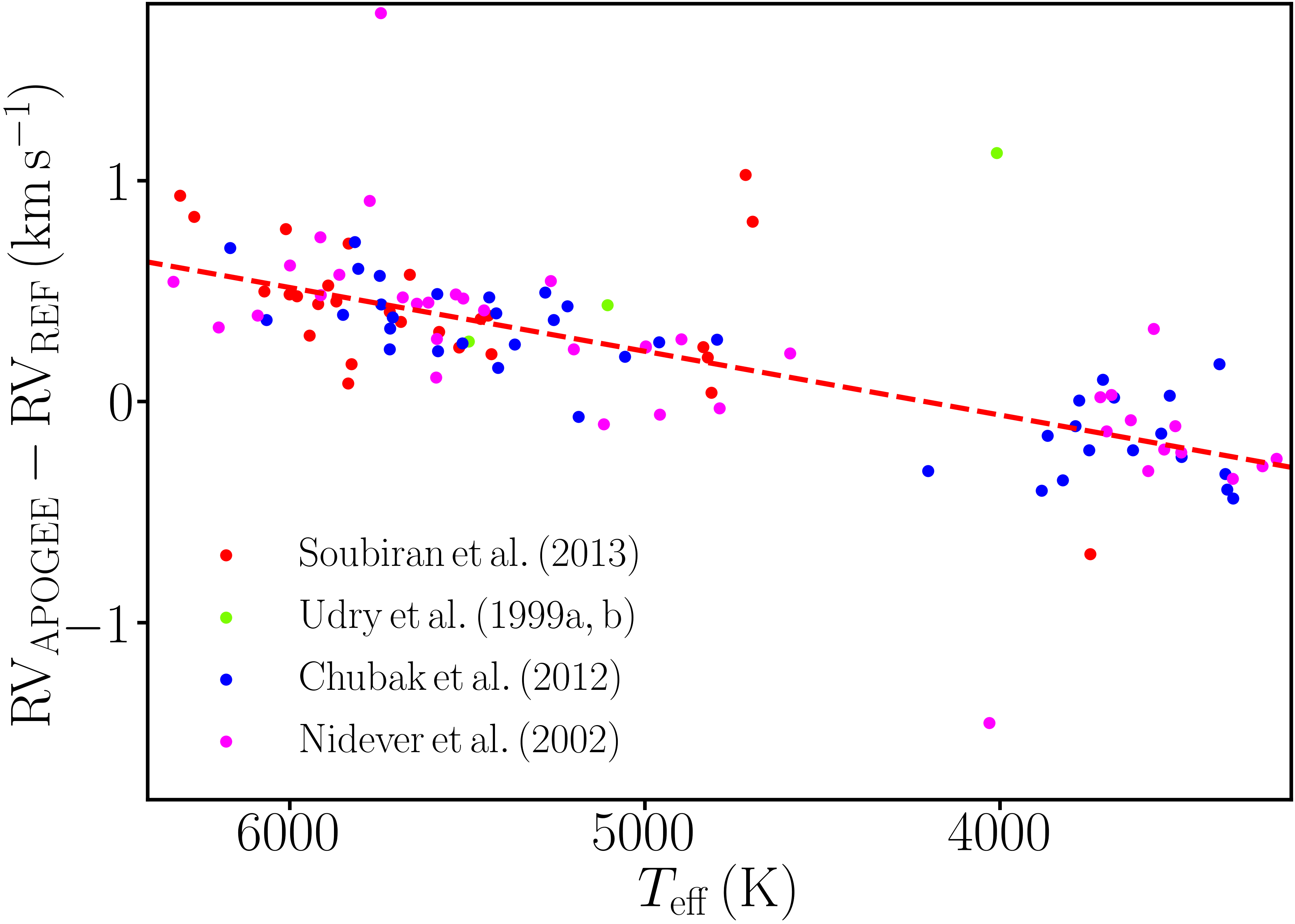}
\caption{The differences between the APOGEE RV ($\rm{RV}_{\rm{APOGEE}}$) and the RV standard star ($\rm{RV}_{\rm{REF}}$) collected from the literature as a function of effective temperature ($T_{\rm{eff}}$). Stars of different colors are reference RV standard stars selected from different catalogs, as marked in the bottom-left corner. The dashed red line is the best linear-fit to the data points.}
\label{fig:RefSTD}
\end{figure}

At present, over tens of thousands of RV standard stars have been defined by various efforts, including about 5000 bright RV standard stars with extreme stability of 15\,m\,s$^{-1}$ over an average baseline of six years constructed by a long monitoring project \citep{1999ASPC..185..383U,1999ASPC..185..367U,2007sf2a.conf..459C,2010A&A...524A..10C,2013A&A...552A..64S,2018A&A...616A...7S,2012arXiv1207.6212C} and over 18,000 standard stars with a median stability of 240 m\,s$^{-1}$ over one-year baseline of a large color and magnitude range constructed from the APOGEE DR14 \citep[][hereafter H18]{2018AJ....156...90H}. However, the number spatial density of current RV standard stars is too low that are hard to calibrate the RV zero points (RVZPs) of the RV measurements from future massive spectroscopic surveys. 

This paper is an update of \citetalias{2018AJ....156...90H}. Thanks to the long-term repeated observations and more southern stars observed during SDSS-IV, the number of RV standard stars has been trebled with a much large sky coverage from the APOGEE DR17, compared to the previous version of \citetalias{2018AJ....156...90H}. The paper is structured as follows. In Section \ref{section:RRSFED}, we correct the possible RVZPs of the RV measurements of APOGEE DR17. In Section \ref{section:ARVSS}, we describe the details of selections of RV standard stars from the APOGEE DR17. In Section \ref{section:CRVSSSS}, we use the selected APOGEE RV standard stars to calibrate the RVZPs of the RAVE, GALAH, LAMOST and {\it Gaia} surveys. Finally, we conclude in Section \ref{section:Sum}.

\section{Corrections of APOGEE RV measurements}
\label{section:RRSFED}

As found in \citetalias{2018AJ....156...90H}, the measured RVs of APOGEE surveys exhibit systematic trend as a function of $T_{\rm eff}$. To correct this trend, 1611 reference RV standard stars, collected from the various literatures \citep{1999ASPC..185..383U,1999ASPC..185..367U,2002ApJS..141..503N,2012arXiv1207.6212C,2013A&A...552A..64S}, are adopted to calibrate the RVZP of APOGEE DR14 \citep{2018ApJS..235...42A}. In this paper, we apply these reference RV standard stars to check the RVZP of the APOGEE DR17 \citep{2022ApJS..259...35A}. The details of the compilation of these reference RV standard stars are detailedly described in \citetalias{2018AJ....156...90H}. Generally, the RVs of these 1611 reference stars are required to have stability better than 100 m\,s$^{-1}$ over a baseline at least one year.

\begin{figure*} 
\centering
\includegraphics[scale=0.14,angle=0]{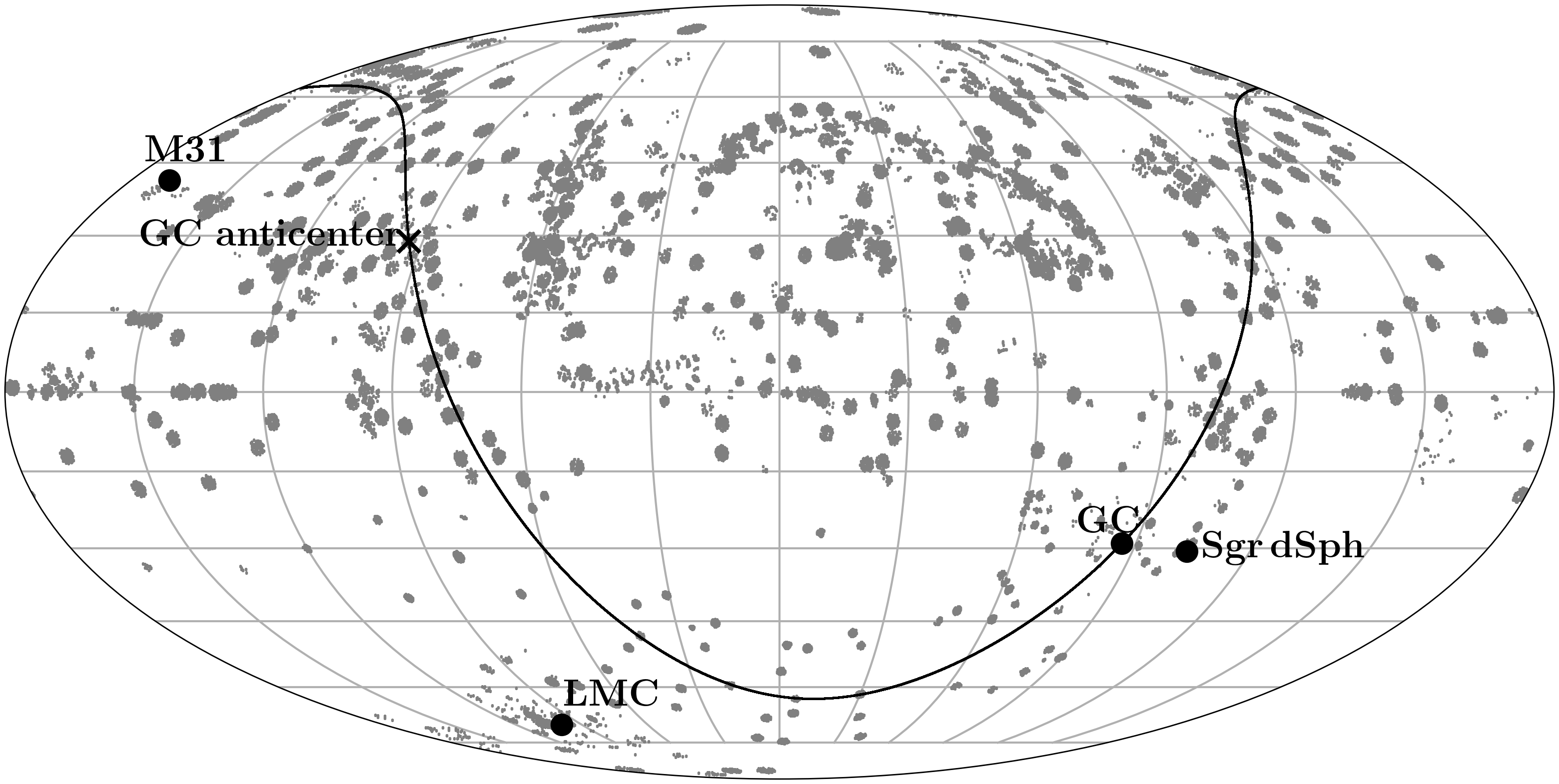}
\caption{Hammer projection in Right Ascension and Declination of the APOGEE RV standard stars. The black line delineates the Galactic plane, on which we mark the positions of the Galactic center (GC) and the Galactic anti-center (GAC). The positions of M31, LMC and Sgr dSph are also marked.}
\label{fig:Fig3}
\end{figure*}

\begin{figure*} 
\centering
\includegraphics[scale=0.14,angle=0]{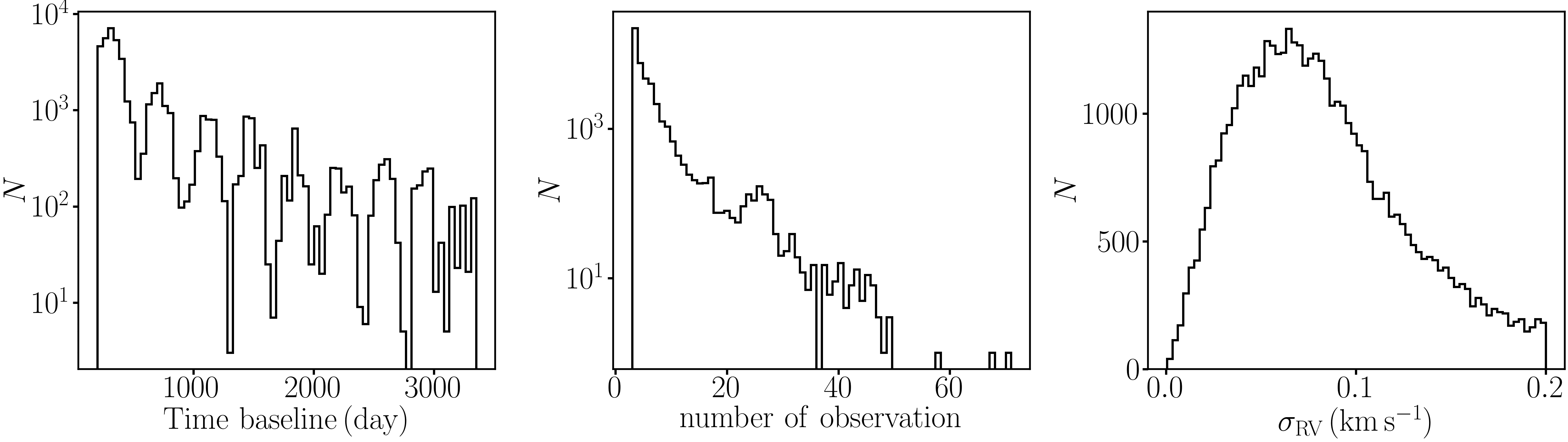}
\caption{The time baseline distribution (left panel), observation number distribution (middle panel) and $\overline{\rm RV}$ weighted standard deviation ($\sigma _{\rm{RV}}$) distribution (right panel) of 46,753 APOGEE RV standard stars.}
\label{fig:Fig4}
\end{figure*}

\begin{figure} 
\centering
\includegraphics[scale=0.14,angle=0]{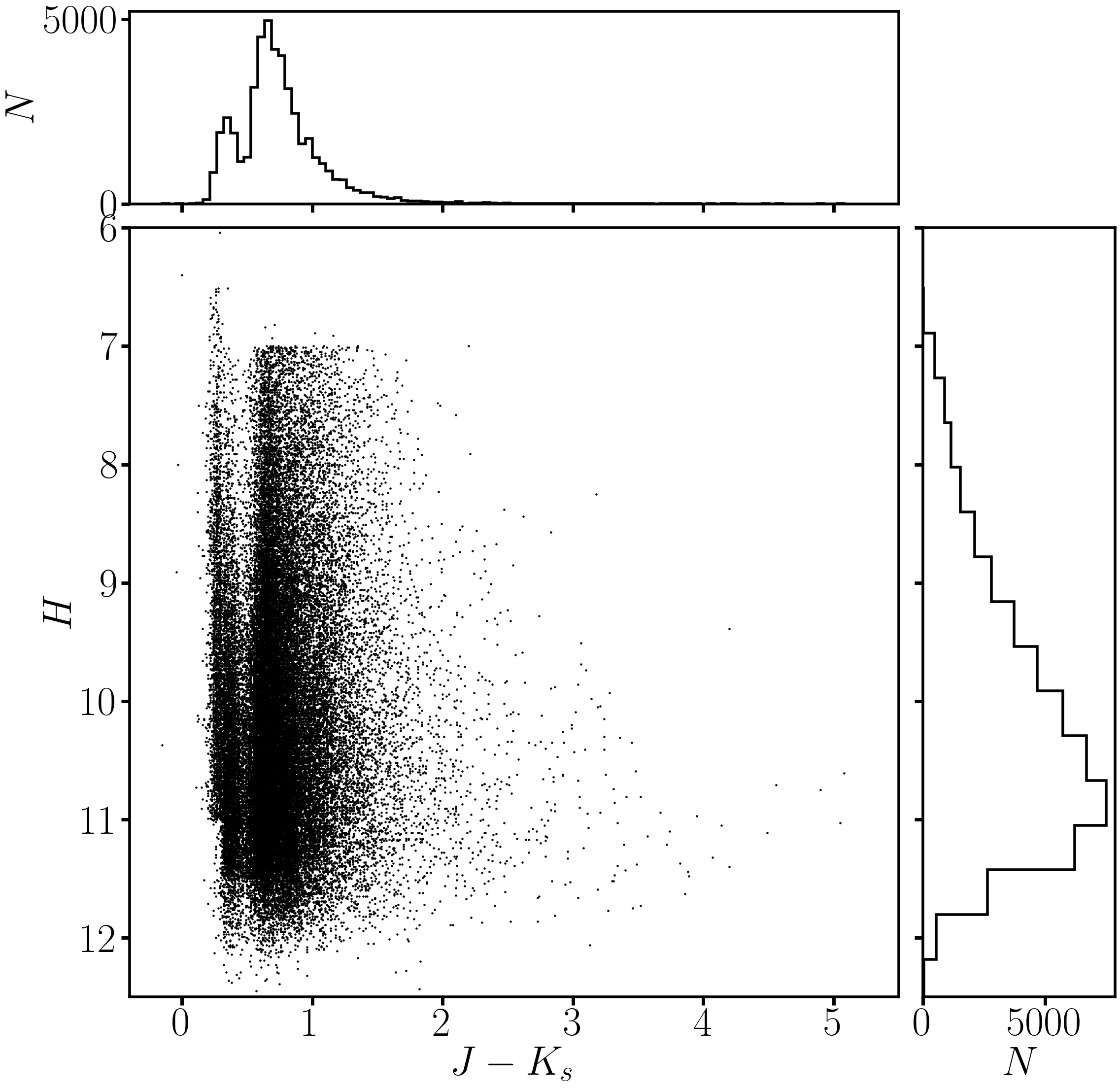}
\caption{The color $J-K_{s}$--magnitude $H$ diagram of APOGEE RV standard stars. The top and right insets are histogram distributions along the $J-K_{s}$ and $H$ axes, respectively.}
\label{fig:Fig5}
\end{figure}

By crossing match the 1611 reference RV standard stars to APOGEE\,DR17, 118 common stars are found to check the RVZPs of APOGEE DR17. 
The RV differences between APOGEE and reference RV standard stars show a significant systematic trend along stellar effective temperature (see Fig.~\ref{fig:RefSTD}).
To describe this trend, a simple linear fit is adopted:
\begin{equation}\label{EQ1}
{\Delta {\rm RV}} = - 1.2146 + 0.2885 \times (T_{\rm eff}/10^{3} {\rm\,K}) {\rm\,km\,s^{-1}}\text{.}
\end{equation}
The coefficients found here are similar to these reported in \citetalias{2018AJ....156...90H}, implying the robustness of the instruments and the RV measurements.

We also cross-matched 4813 RV standard stars constructed by \citet{2018A&A...616A...7S} with APOGEE DR17 and 205 common stars are left. The systematic trend found by these common stars is generally consistent with that shown in Fig.~\ref{fig:RefSTD} for stars with $T_{\rm eff} > 4000$\,K. For cold stars with $T_{\rm ff} < 4000$\,K, few stars are found.

\begin{table*}[!t]
\caption{The final sample of APOGEE radial velocity standard stars}
\label{table:RV-STD}
\resizebox{17cm}{4.4cm}{
\begin{threeparttable}\centering
\begin{tabular}{ccccccccccc}
\cline{1-11}
\multirow{1}{*}   Name & $H$ & $J-K_{\rm s}$ & $T_{\rm eff}$ & $\overline{\rm RV}$ & $I_{\rm ERV}$ & $\sigma_{\rm RV}$ & $n$ & $\sigma_{\overline{\rm RV}}$ & $\Delta T$ & Mean MJD \\
&&&(K)&(km\,s$^{-1}$)&(km\,s$^{-1}$)&(km\,s$^{-1}$)&&(km\,s$^{-1})$&(days)&($-50000$)\\
\cline{1-11}
J00:00:00.20-19:24:49.9 & 10.74 & 0.40 & 5501 & 18.698 & 0.0223 & 0.0926 & 3 & 0.0535 & 1091 & 7904 \\
J00:00:00.32+57:37:10.3 & 10.64 & 0.42 & 6162 & -21.057 & 0.0261 & 0.1424 & 6 & 0.0581 & 776 & 6328 \\
J00:00:00.68+57:10:23.4 & 10.13 & 0.65 & 5031 & -12.914 & 0.0165 & 0.1284 & 3 & 0.0742 & 776 & 6389 \\
J00:00:05.07+56:56:35.3 & 8.72 & 0.77 & 4981 & 4.348 & 0.0114 & 0.1125 & 5 & 0.0503 & 421 & 8651 \\
J00:00:05.35+15:04:34.4 & 11.17 & 0.59 & 4914 & 18.229 & 0.0149 & 0.0275 & 3 & 0.0159 & 2493 & 7416 \\
J00:00:12.11-19:03:38.3 & 10.53 & 0.36 & 5740 & 23.782 & 0.0250 & 0.0727 & 3 & 0.0420 & 1091 & 7904 \\
J00:00:12.17-19:49:30.6 & 10.77 & 0.39 & 5645 & -5.845 & 0.0192 & 0.0427 & 3 & 0.0247 & 1091 & 7904 \\
J00:00:12.43+55:24:39.1 & 11.48 & 0.84 & 4659 & -114.291 & 0.0156 & 0.0367 & 3 & 0.0212 & 2954 & 6813 \\
J00:00:13.62-19:13:04.2 & 11.06 & 0.37 & 5555 & -30.755 & 0.0223 & 0.0585 & 3 & 0.0338 & 1091 & 7904 \\
J00:00:20.05+57:03:46.8 & 9.79 & 1.06 & 4355 & -52.964 & 0.0080 & 0.0789 & 8 & 0.0279 & 2972 & 7296 \\
...&...&...&...&...&...&...&...&...&...&...\\
J17:15:48.00-30:18:23.9 & 8.73 & 1.60 & 3392 & -55.954 & 0.0150 & 0.0793 & 3 & 0.0458 & 327 & 8493 \\
J17:15:48.77+59:10:10.1 & 11.05 & 0.60 & 4832 & -18.619 & 0.0207 & 0.0211 & 4 & 0.0106 & 844 & 6958 \\
J17:15:49.98+42:38:09.7 & 10.85 & 0.61 & 4887 & 11.785 & 0.0170 & 0.0514 & 3 & 0.0297 & 324 & 6143 \\
J17:15:50.95-42:57:40.5 & 10.18 & 1.16 & 4596 & -51.058 & 0.0176 & 0.0176 & 3 & 0.0102 & 355 & 8582 \\
J17:15:51.31+23:55:24.4 & 10.26 & 0.59 & 4949 & 19.250 & 0.0162 & 0.0544 & 3 & 0.0314 & 314 & 6521 \\
J17:15:51.77+30:59:06.2 & 10.80 & 0.65 & 4708 & -67.426 & 0.0098 & 0.1100 & 8 & 0.0389 & 1358 & 7726 \\
J17:15:52.42+65:42:59.0 & 10.19 & 0.20 & 6162 & -21.094 & 0.0266 & 0.1295 & 4 & 0.0648 & 391 & 8469 \\
J17:15:52.48+56:43:37.6 & 9.15 & 0.74 & 4565 & 4.726 & 0.0081 & 0.1234 & 11 & 0.0372 & 1183 & 8446 \\
J17:15:52.74+29:07:36.8 & 11.25 & 0.84 & 4455 & -81.656 & 0.0221 & 0.1205 & 4 & 0.0603 & 455 & 7856 \\
J17:15:52.99+30:52:49.0 & 8.90 & 0.23 & 6151 & -6.513 & 0.0152 & 0.1330 & 10 & 0.0420 & 1359 & 7740 \\
...&...&...&...&...&...&...&...&...&...&...\\
\cline{1-11}
\end{tabular}
\end{threeparttable}}
\end{table*}

\section{APOGEE Radial velocity standard stars}
\label{section:ARVSS}
\subsection{APOGEE Survey} 
\label{section:APOGEESurvey}

\begin{figure}[!b]
\centering
\includegraphics[scale=0.09,angle=0]{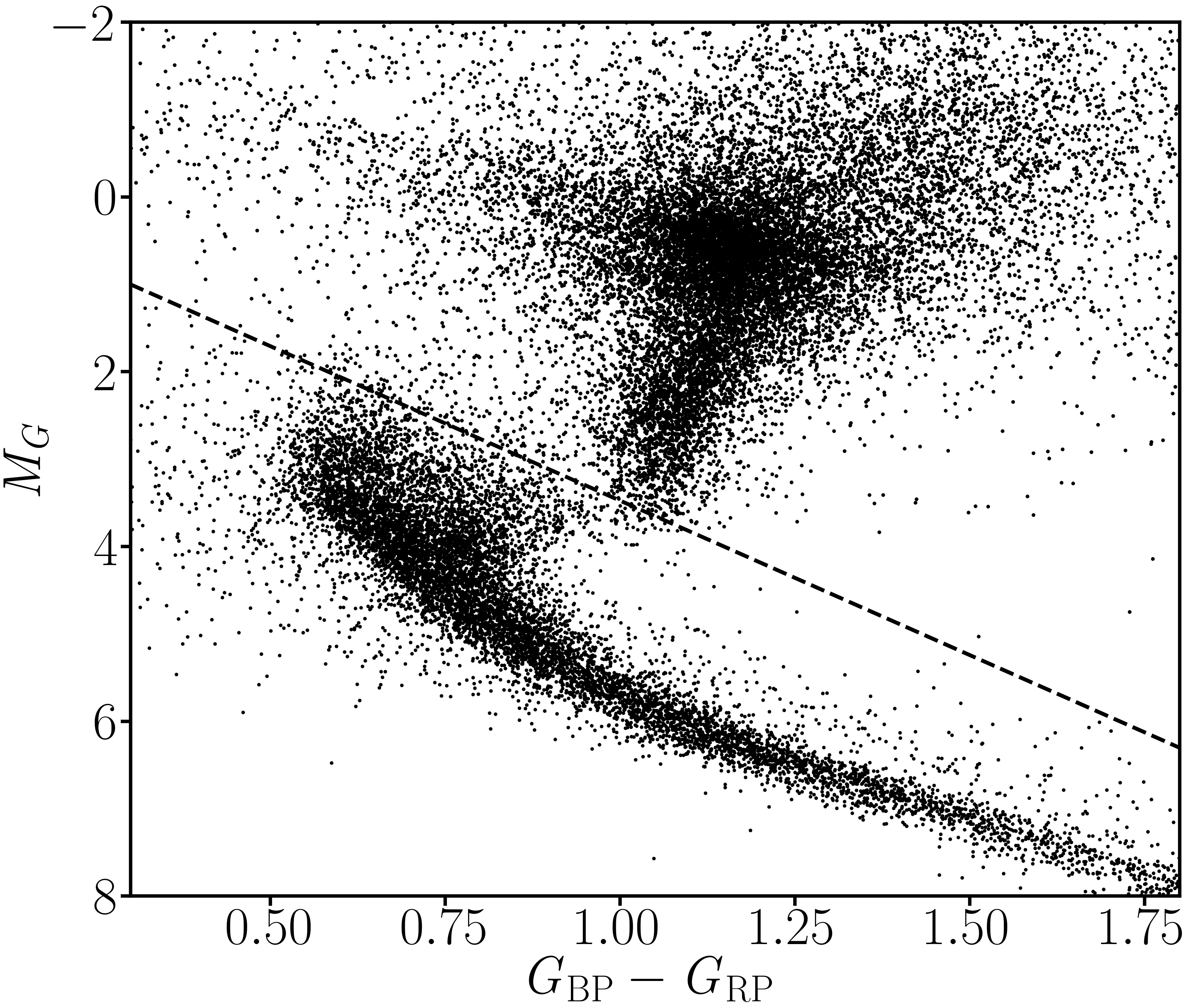}
\caption{The color $G_{\rm{BP}}-G_{\rm{RP}}$--magnitude $M_{G}$ distribution of APOGEE RV standard stars. The black dotted line is the empirical line to separate giants and main-sequence dwarf stars.}
\label{fig:CMSTD}
\end{figure}

APOGEE \citep{2013AJ....146...81Z,2016AN....337..863M,2017AJ....154...94M,2017AJ....154..198Z} is a large-scale high-resolution ($R\sim 22,500$) spectroscopic survey in the near-infrared ($H$-band $1.51-1.70$\,$\mu\rm{m}$) part, provided by the 2.5-meter Sloan Foundation Telescope \citep{2006AJ....131.2332G} and the 1-meter NMSU Telescope \citep{2010AdAst2010E..46H} at the Apache Point Observatory (APO) in the Northern Hemisphere, and the 2.5-meter Ir{\'e}n{\'e}e du Pont Telescope \citep{1973ApOpt..12.1430B} at the Las Campanas Observatory (LCO) in the Southern Hemisphere. The APOGEE survey is an important part of the SDSS-III \citep{2011AJ....142...72E} and SDSS-IV \citep{2017AJ....154...28B} programs. The APOGEE survey is called ``APOGEE" or ``APOGEE-1" in SDSS-III, and it is called ``APOGEE-2" in SDSS-IV. APOGEE-1 started its data collection in 2011 and ended in 2014. The SDSS DR10 publicly released the APOGEE-1 three-year dataset, which was subsequently followed by two additional releases in 2015 and 2016. This accomplishment successfully fulfilled the stated objective of observing over 100,000 stars with a limiting magnitude of $H=12.2$ mag and spectral signal-to-noise ratio (SNR) greater than 100. APOGEE-2 is a constituent program of the SDSS-IV initiative, which commenced in 2014 and finished in 2021. In addition to collect the data in the Northern Hemisphere, APOGEE-2 also adopted the 2.5-meter Ir{\'e}n{\'e}e du Pont Telescope mounted on the LCO to expand the observation sky coverage to the Southern Hemisphere. The recently released SDSS DR17 \citep{2022ApJS..259...35A} includes the latest version of the APOGEE survey, including more than 657,000 stars, and this version is also the final version of all APOGEE-1 and APOGEE-2 data. The measurement of RV has an uncertainty of $100$\,m\,s$^{-1}$ and a zero point offset of $500$\,m\,s$^{-1}$ \citep{2015AJ....150..173N}. Typical uncertainties for $T_{\rm{eff}}$, $\rm{log}\,\it{g}$, and [Fe/H] are better than $150$\,K, $0.2$\,dex, and $0.1$\,dex, respectively \citep{2013AJ....146..133M,2016AJ....151..144G}.

\begin{table*}[!t]
\caption{Comparisons of RVs yielded by those of the APOGEE RV standard stars and RAVE, LAMOST, GALAH and {\it Gaia} surveys}
\label{table:Com-RV-STD}
\resizebox{\textwidth}{!}{ 
\begin{threeparttable}
\begin{tabular}{lccccl}
\hline
Source & $\Delta {\rm RV}$ (km\,s$^{-1}$) & s.d. (km\,s$^{-1}$) & $\overline{\rm SNR}$& $N$ & calibration relationship\\
\hline
RAVE&$+0.149$&$1.358$&$56$&$1284$ &  $\Delta \rm{RV}\,\,\,\,\,\,\,\,=\,\,\,\,+0.1058 + 0.4175\times [Fe/H]$, for $-1<$\,[Fe/H]\,$<0.5$\\
LAMOST LRS&$+4.574$&$3.844$&$96$&$15600$ & $\Delta \rm{RV}\,\,\,\,\,\,\,\,=\,\,\,\,+4.57$\\
LAMOST MRS&$-0.006$&$1.053$&$61$&$6431$ & $\Delta \rm{RV}\,\,\,\,\,\,\,\,=\,\,\,\,-0.2233-0.1772 \times \left[(T_{\rm{eff}}-3900)/10^{3}\right]+1.3234 \times \left[(T_{\rm{eff}}-3900)/10^{3}\right]^{2}-1.2789 \times$\\
 & & & & & $\,\,\,\,\,\,\,\,\,\,\,\,\,\,\,\,\,\,\,\,\,\,\,\,\,\,\,\,\,\,\,\,\, \left[(T_{\rm{eff}}-3900)/10^{3}\right]^{3}+0.3138 \times \left[(T_{\rm{eff}}-3900)/10^{3}\right]^{4}$, for $3900<T_{\rm{eff}}<6500$\\
GALAH&$-0.031$&$0.299$&$40$&$1839$ & $\Delta \rm{RV}\,\,\,\,\,\,\,\,=\,\,\,\,-0.1834+1.3138\times \rm{log}\,{\it g} - 1.9784\times \rm{log}\,{\it g}^{2} + 1.3765\times \rm{log}\,{\it g}^{3} -0.4979\times \rm{log}\,{\it g}^{4}+ $ \\
 & & & & & $\,\,\,\,\,\,\,\,\,\,\,\,\,\,\,\,\,\,\,\,\,\,\,\,\,\,\,\,\,\,\,\,\,0.0879\times\rm{log}\,{\it g}^{5} - 0.0059\times \rm{log}\,{\it g}^{6}$, for $0.8<$\,log\,$g<4.7$\\
{\it Gaia}& {$+0.014$}& {$0.561$}&{--}&{$43214$} & $\Delta \rm{RV}\,\,\,\,\,\,\,\,=\,\,\,\,0.0163-0.0335\times (G_{\rm{RVS}}-8.2)+ 0.1327\times (G_{\rm{RVS}}-8.2)^2-0.0930\times (G_{\rm{RVS}}-8.2)^3$ \\
& & & & & $\,\,\,\,\,\,\,\,\,\,\,\,\,\,\,\,\,\,\,\,\,\,\,\,\,\,\,\,\,\,\,\,\,+0.0232\times (G_{\rm{RVS}}-8.2)^4-0.0020\times(G_{\rm{RVS}}-8.2)^5$, for $8.2<G_{\rm{RVS}}<14$\\
& & & & & $\Delta \rm{RV}^{\prime}$\tnote{(a)}$\,\,\,\,\,\,\,\,=\,\,\,\,-0.7751+1.0987\times (G_{\rm{BP}}-G_{\rm{RP}})-0.4549\times (G_{\rm{BP}}-G_{\rm{RP}})^{2}+0.0450\times (G_{\rm{BP}}-G_{\rm{RP}})^{3}$  \\
& & & & & $\,\,\,\,\,\,\,\,\,\,\,\,\,\,\,\,\,\,\,\,\,\,\,\,\,\,\,\,\,\,\,\,\,+0.0064\times (G_{\rm{BP}}-G_{\rm{RP}})^4$, for $0.5<G_{\rm{BP}}-G_{\rm{RP}}<3.3$\\
\hline
\end{tabular}
\begin{tablenotes}
\item[$^a$] Here $\Delta$\,RV$^{\prime}$ denotes the RV differences after the corrections of  $G_{\rm{RVS}}$ dependent systematics.
\end{tablenotes}
\end{threeparttable}}
\end{table*}

\subsection{Selecting RV standard stars from the APOGEE DR17}

Following \citetalias{2018AJ....156...90H}, to select RV standard stars from APOGEE DR17, we defined the weighted mean RV ($\overline{\rm RV}$), internal error of  $\overline{\rm RV}$ ($I_{\rm ERV}$), $\overline{\rm RV}$ wighted standard deviation ($\sigma_{\rm RV}^{2}$), and uncertainty of $\overline{\rm RV}$ ($\sigma_{\overline{\rm RV}}$) for each star separately:
\begin{enumerate}[label=\arabic*)]
\item $\overline{\rm RV} = \frac{{\sum\limits_{i=1}^{n}\rm RV_{i}}w_{i}}{\sum\limits_{i = 1}^{n}w_{i}}$, where $w_{i}$ is the weight assigned by the individual RV measurement error $\epsilon_{i}$, that is, $1/\epsilon_{i}^{2}$, $n$ is the total number of observations;
\item $I_{\rm ERV} = \frac{{\sum\limits_{i=1}^{n}\rm \epsilon_{i}}w_{i}}{\sum\limits_{i = 1}^{n}w_{i}}$;
\item $\sigma_{\rm RV}^{2} = \frac{\sum\limits_{i=1}^{n}w_{i}}{(\sum\limits_{i=1}^{n}w_{i})^{2}-\sum\limits_{i=1}^{n}w_{i}^{2}}\sum\limits_{i}^{n}w_{i}({\rm RV_{i}} - \overline{\rm RV})^{2}$;
 
 \item $\sigma_{\overline{\rm RV}}=$ max($\sigma_{\rm RV}/\sqrt{N}$, $I_{\rm RV}/\sqrt{N}$).
\end{enumerate}
We utilize the symbol $\Delta T$ to denote the time baseline and MJD to represent the mean modified julian day of the $n$ observations. In the following step, we selected RV standard stars based on the following criteria: $\Delta T >200$\,days, $n\geq3$, $\rm{SNR}_{\rm low}\geq50$ and $\sigma_{\rm RV}\leq 200$\,m\,s$^{-1}$, where $\rm{SNR}_{\rm low}$ represents the  lowest SNR for the multiple spectroscopic visits for each star.

Through the above cuts, a total of 46,753 APOGEE RV standard stars were selected. The spatial distribution is shown in Fig.~\ref{fig:Fig3}, with a full sky coverage. The distributions of time baseline $\Delta T$, number of observations $n$ and $\sigma_{\rm RV}$ for these RV standard stars  are shown in Fig.~\ref{fig:Fig4}. Their $\Delta T$ are all greater than 200\,days (54\% longer than one year and 10\% longer than five years). The average number of observations for these stars is 5. The median $\sigma_{\rm RV}$ of all standard stars is $71.75$\,m\,s$^{-1}$, corresponding to a median stability ($3\sigma_{\rm RV}$) of $215.25$ \,m\,s$^{-1}$, better than $240$\,m\,s$^{-1}$ of the \citetalias{2018AJ....156...90H} sample. We show the color-(absolute) magnitude distributions of these stars, see Fig.~\ref{fig:Fig5} for $H$ against $J-K_{s}$, and Fig.~\ref{fig:CMSTD} for absolute $M_{G}$ against $G_{\rm{BP}}-G_{\rm{RP}}$. Due to the selection effect of the APOGEE survey, 81\% of RV standard stars are redder than 0.5 on the $J-K_{s}$. The Fig.~\ref{fig:CMSTD} contains 30,268 RV standard stars with measurable distances, all of which have been corrected for interstellar extinction using the 2D dust map from \citet{1998ApJ...500..525S}. We employ the empirical relation $M_{G}=3.53\times (G_{\rm{BP}}-G_{\rm{RP}})-0.06$ to distinguish giants from main-sequence dwarf stars (see dashed line). Amongst them, 62\% are red giants and 38\% are main-sequence dwarf stars. We list 46,753 APOGEE RV standard stars including name, $H$, $J-K_{s}$, $T_{\rm{eff}}$, $\overline{\rm RV}$ (after RVZP correction by Equation~\ref{EQ1}), $I_{\rm{ERV}}$, $\sigma_{\rm{RV}}$, $n$, $\sigma_{\overline{\rm RV}}$, $\Delta T$ and mean MJD information in Table~\ref{table:RV-STD}.

\begin{figure*} 
\centering
\includegraphics[scale=0.13,angle=0]{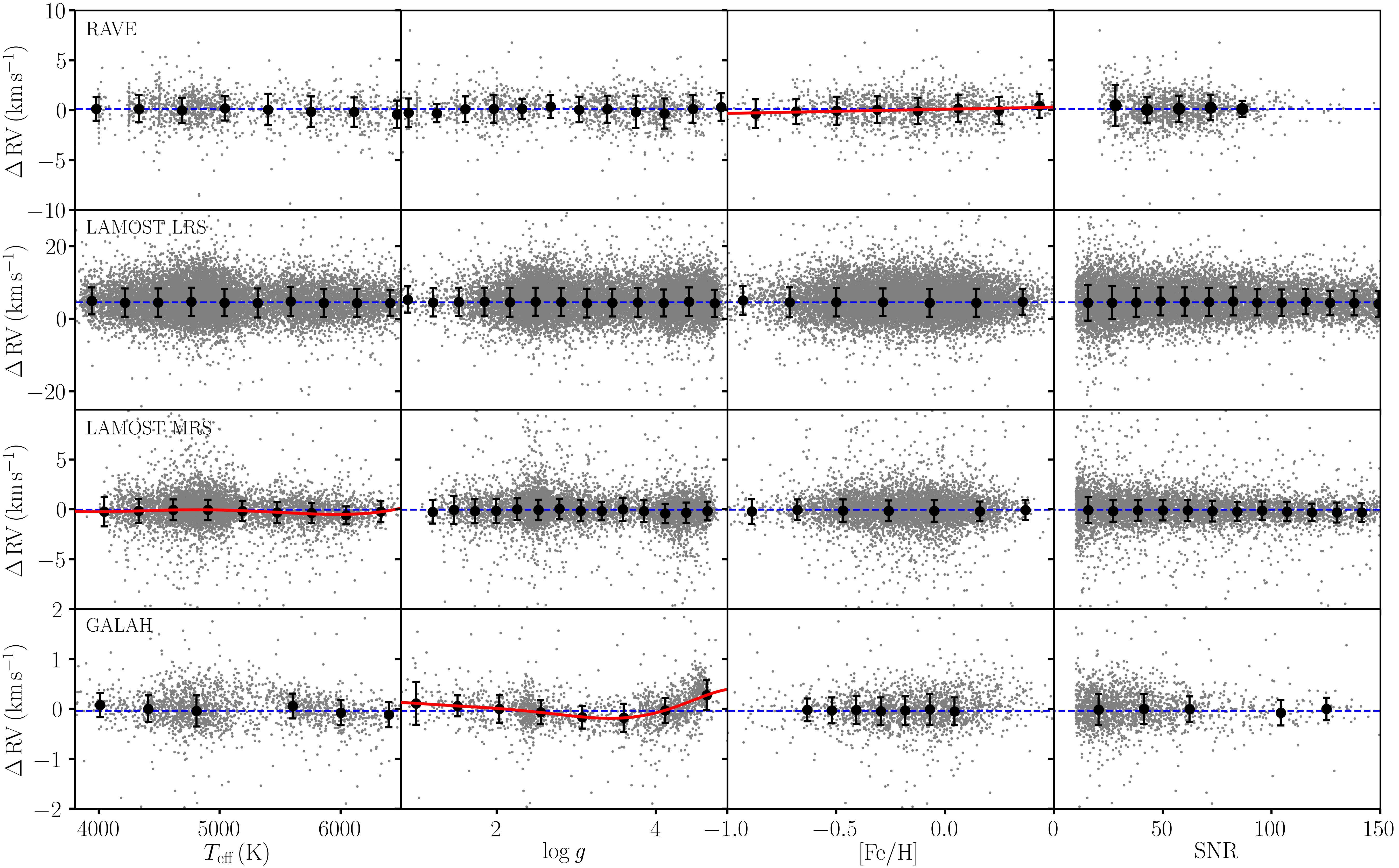}
\caption{$\rm{RV}$ differences of common stars (gray dots) between APOGEE RV standard stars and different surveys (top panel: RAVE; middle two panels: LAMOST LRS and MRS (zero-point corrected RV measurements after October 19, 2018); bottom panel: GALAH) as a function of $T_{\rm{eff}}$ (first column), log\,$\it{g}$ (second column), [Fe/H] (third column) and SNR (fourth column). The black dots and their error bars in each subpanel represent the mean and standard deviation of $\Delta$\,RV in bins of individual parameters ($T_{\rm{eff}}$, log\,$\it{g}$, [Fe/H] and SNR). Blue dashed lines represent the mean difference as shown in Table~\ref{table:Com-RV-STD}. Red lines present polynomial fits of the RV difference as functions of stellar parameters (the coefficients are given in Table~\ref{table:Com-RV-STD}).}
\label{fig:Fig6}
\end{figure*}

\section{Calibrations of radial velocity scales for large-scale stellar spectroscopic surveys}
\label{section:CRVSSSS}

\begin{figure*} 
\centering
\includegraphics[scale=0.13,angle=0]{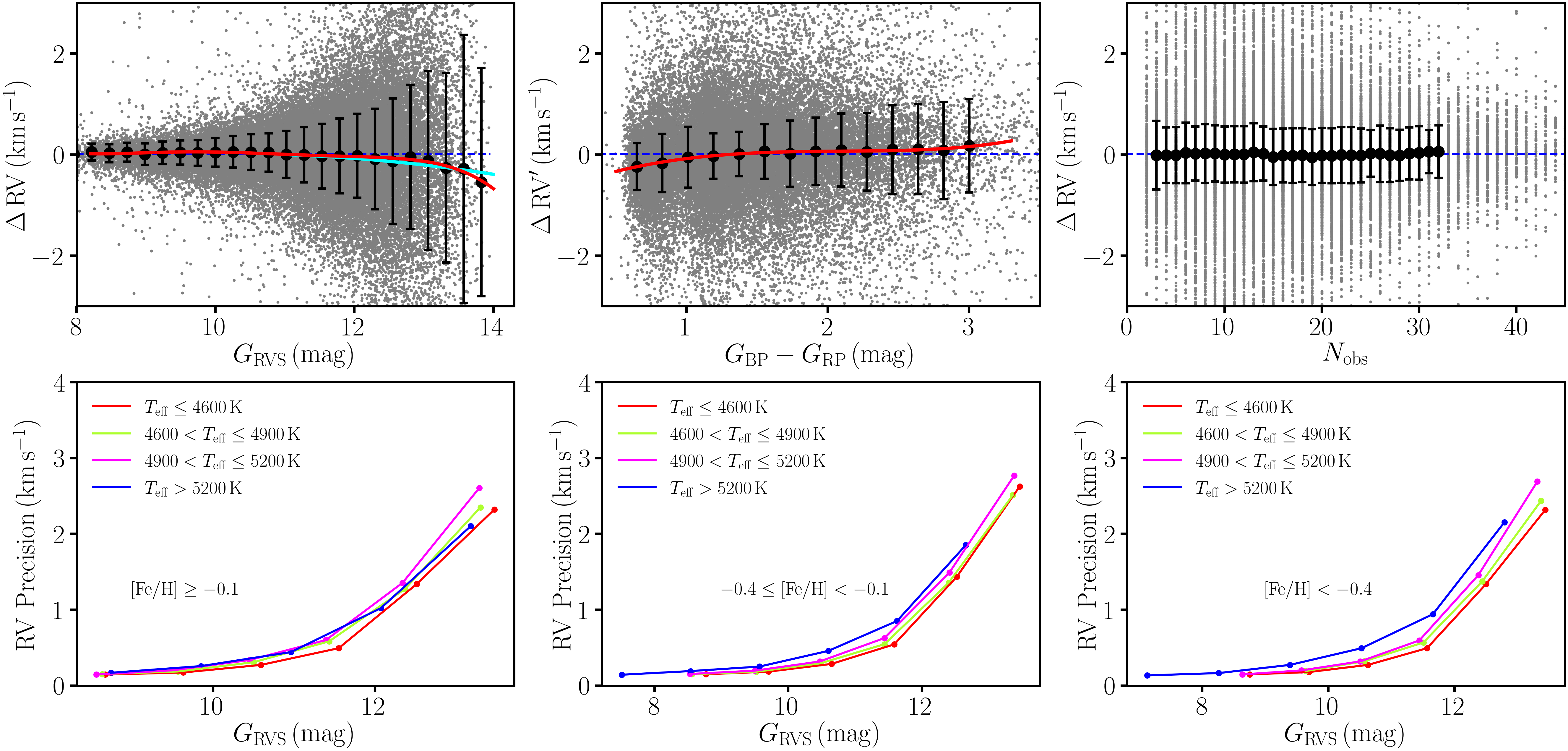}
\caption{{\it Top panel}: Similar to Fig.~\ref{fig:Fig6}, but this figure presents common stars between APOGEE RV standard stars and {\it Gaia} DR3, those subplots show systematic trend of $\Delta \rm{RV}$ with $G_{\rm RVS}$ (left), $G_{\rm{BP}}-G_{\rm{RP}}$ (middle), and $N_{\rm{obs}}$ (right). Red lines represents polynomial fits of $\Delta$\,RV with $G_{\rm{RVS}}$ or $G_{\rm{BP}}-G_{\rm{RP}}$, and the coefficients are listed in Table~\ref{table:Com-RV-STD}. $\Delta$\,RV$^{\prime}$ in the middle panel represents the RV differences after the corrections of $G_{\rm RVS}$ dependent systematics. Cyan line is a second-order polynomial taken from \citet{2022arXiv220605902K} to describe the systematic trend of $\Delta \rm{RV}$ with $G_{\rm{RVS}}$ when $G_{\rm{RVS}}>11$\,mag. {\it Bottom panel}: The RV precision of {\it Gaia} DR3, as a function of $G_{\rm{RVS}}$, is reported through comparison with APOGEE RV standard stars, corresponding to metal-rich stars ([Fe/H] $\geq -0.1$, left), middle metallicity stars ($-0.4\leq $ [Fe/H] $<-0.1$, middle), and relatively metal-poor stars ([Fe/H] $< -0.4$, right). The different colored curves are calculated for each effective temperature range of the APOGEE RV standard stars as labeled in the top-left corner on each panel.}
\label{fig:Fig7}
\end{figure*}

Next, we use these 46,753 APOGEE RV standard stars to check the RVZPs of the RAVE, LAMOST, GALAH, and {\it Gaia} surveys. The calibration results are shown in Figs.~\ref{fig:Fig6}, \ref{fig:Fig7} and Table~\ref{table:Com-RV-STD}.

\textbf{(i) The RAVE survey:} The RAVE survey has collected 520,781 medium-resolution ($R \sim 7500$) spectra centered on the Ca~I triplet (8410--8795\AA) range. The survey has released 457,588 individual stars randomly selected from the Southern Hemisphere stars with $9 < I < 12$ using the multi-object spectrograph 6dF on the Australian Astronomical Observatory's 1.2m UK Schmidt Telescope. Estimations of RV, atmospheric parameters ($T_{\rm eff}$, $\rm{log}\,\it{g}$ and [Fe/H]), and $\alpha$ element abundances were described in \citet{2017AJ....153...75K}.

We cross-matched RAVE DR5 with our 46,753 APOGEE RV standard stars, resulting in a total of 1284 common stars with $\rm{SNR}>10$. The comparisons show a mean $\Delta \rm{RV}$ (APOGEE RVs minus RAVE) of $+0.149$\,km\,s$^{-1}$, with a standard deviation of $1.358$\,km\,s$^{-1}$. We show the systematic trends of $\Delta \rm{RV}$ with $T_{\rm eff}$, log\,$g$, [Fe/H] and SNR in Fig.~\ref{fig:Fig6}. There is no obvious systematic trend of $\Delta \rm{RV}$ with $T_{\rm eff}$, log\,$g$ and SNR, however, $\Delta \rm{RV}$ shows a weak linear trend with [Fe/H]. This trend can be described by $\Delta \rm{RV}=0.1058 + 0.4175\times [Fe/H]$ (see Table~\ref{table:Com-RV-STD}).

\textbf{(ii) The LAMOST survey:} LAMOST is a 4-meter quasi-meridian reflecting Schmidt telescope \citep{2012RAA....12.1197C}. The telescope is equipped with 4000 fibers distributed in a field of view with a diameter of $5^{\circ}$. Within one exposure, LAMOST can obtain 4000 optical low-resolution spectra (LRS; R$\sim $2000; with wavelength coverage between 3700 and 9000\AA) or medium-resolution spectra (MRS; R$\sim $7500; with two wavelength windows of 4950-5350\AA\,\,and 6300-6800\AA , respectively). 

Over ten million LRS spectra has been released in the recent DR9 of LAMOST (\url{http://www.lamost.org/dr9/v1.1/}). A total of 7,060,436 stars in the AFGK Stellar Parameters catalog of LAMOST DR9 LRS have measurements of RV and stellar atmospheric parameters, which are derived by the official stellar parameter pipeline: LAMOST Stellar Parameter Pipeline \citep[LASP;][]{2015RAA....15.1095L}. To check the RVZPs of the LAMOST LRS RVs, we cross-matched the LAMOST DR9 AFGK Stellar Parameters catalog with the RV standard stars. 15,600 common stars are found with $\rm{SNR}>10$ (average $\overline{\rm{SNR}}=96$, see Table~\ref{table:Com-RV-STD}). The comparisons show a mean $\Delta \rm{RV}$ (APOGEE RVs minus LAMOST) of $+4.574$\,km\,s$^{-1}$ and a scatter of 3.844\,km\,s$^{-1}$. No obvious systematic trend with $T_{\rm eff}$, log\,$g$, [Fe/H] and SNR are detected for LAMOST LRS RVs (see middle panels of Fig.~\ref{fig:Fig6}). 

The MRS parameter catalog released in LAMOST DR9 contains measurements of stellar atmospheric parameters and RV for over 1.6 million stars from 8 million MRS spectra. To check the RVZPs of the LAMOST MRS RVs, we cross-matched the LAMOST DR9 MRS parameter catalog with the RV standard stars, resulting in 6,431 common stars with $\rm{SNR}>10$ (see Table~\ref{table:Com-RV-STD}). By comparing their RVs (measurements from LASP) with the standard stars, multiple peaks are found in the RV difference distribution. These peaks are dominated by two mainnes, one occurring before October 19, 2018 (MJD$=$58,410) and another after this date. Prior to October 19, 2018, the mean $\Delta \rm{RV}$ (APOGEE RVs minus LAMOST) was 6.843\,km\,s$^{-1}$ with a standard deviation of 1.202\,km\,s$^{-1}$, while after that date, the mean $\Delta \rm{RV}$ was 0.727\,km\,s$^{-1}$ with a standard deviation of 1.183\,km\,s$^{-1}$. The main reason for such a significant transition in mean $\Delta \rm{RV}$ arises from the use of different wavelength calibration lamps. Prior to October 19, 2018, the Sc lamp was employed to calibrate the wavelength of the LAMOST test observation spectra, whereas the Th--Ar lamp has been used since then \citep{2019ApJS..244...27W,2021ApJS..256...14Z}. LAMOST MRS provides zero-point corrected RV measurements, with the aforementioned offsets largely corrected. If considering the formal survey started from October 19, 2018, the offset-corrected RVs from LAMOST MRS agree very well with those of the APOGEE RV standard stars, with a nil zero-point and a scatter of 1.05\,km\,s$^{-1}$ (see middle panel of Fig.~\ref{fig:Fig6}). However, the mean RV differences still show a systematic trend with $T_{\rm{eff}}$ (see middle panels of Fig.~\ref{fig:Fig6}). This trend can be described by a fourth-order polynomial in $T_{\rm{eff}}$ (see Table~\ref{table:Com-RV-STD}).

\textbf{(iii) The GALAH survey:} The GALAH survey is a large-scale stellar spectroscopic survey. The aim is to collect high-resolution ($R = $\,28,000) spectra of approximately one million stars in the optical band (four discrete optical wavelength ranges: 4713--4903\AA , 5648--5873\AA , 6478--6737\AA , and 7585--7887\AA) using the HERMES spectrograph installed on the 3.9\,m Anglo-Australian Telescope (AAT) at the Siding Spring Observatory \citep{2015MNRAS.449.2604D}. In the third data release \citep[DR3,][]{2021MNRAS.506..150B}, GALAH provided a total of 678,423 spectra of 588,571 unique stars, including measurements RVs, stellar atmospheric parameters and individual element abundances.

We cross-matched the APOGEE RV standard stars with GALAH DR3 to examine the RVZP of the GALAH RVs. A total of 1839 common stars with $\rm{SNR}>10$ were found, with average $\overline{\rm{SNR}}$ of 40 (Table~\ref{table:Com-RV-STD}). The mean value and standard deviation of the $\Delta \rm{RV}$ (APOGEE RVs minus GALAH) are $-0.031$\,km\,s$^{-1}$ and $0.299$\,km\,s$^{-1}$, respectively. Fig.~\ref{fig:Fig6} (bottom panel) shows the systematic trends of $\Delta \rm{RV}$ with $T_{\rm eff}$, log\,$g$, [Fe/H] and SNR. It can be seen from the plot that $\Delta \rm{RV}$ has no trend with [Fe/H] and SNR. However, $\Delta \rm{RV}$ exhibited curved trend with both $T_{\rm eff}$ and log\,$g$. Through our validation, we find that the systematic trend of GALAH RVs are dominated by log\,$g$. The trend can be described by a sixth-order polynomial about log\,$g$, which the coefficients are presented in Table~\ref{table:Com-RV-STD}.

\textbf{(iv) The Gaia survey:} The European Space Agency (ESA) satellite {\it Gaia} \citep{2016A&A...595A...1G} recently released the Data Release 3 \citep[DR3;][]{2022arXiv220800211G}, which provides astrometric and photometric data for more than 1.8 billion sources. Compared with {\it Gaia} DR2, {\it Gaia} DR3 provides more than 33 million stars with measurements of $\rm{RV}$ \citep{2022arXiv220605902K} and more than 470 million stars with measurements of atmospheric parameters \citep{2022arXiv220605992F}. The median value of RV measurement accuracy is $1.3$\,km\,s$^{-1}$ at $G_{\rm{RVS}}=12$\,mag and $6.4$\,km\,s$^{-1}$ at $G_{\rm{RVS}}=14$\,mag. The RVZP of the {\it Gaia} DR2 has a systematic trend with $G_{\rm{RVS}}$, which shows $\Delta \rm{RV} = 0$\,km\,s$^{-1}$ at $G_{\rm{RVS}}=11$\,mag and $\Delta \rm{RV}=0.40$\,km\,s$^{-1}$ at $G_{\rm{RVS}}=14$\,mag \citep{2018AJ....156...90H,2022arXiv220605902K}.

To check the RVZP of the {\it Gaia} DR3, we cross-matched our APOGEE RV standard stars with {\it Gaia} DR3 to obtain 43,214 common stars with $3200$\,K$<T_{\rm{eff}}<6400$\,K and $\rm{SNR}>5$. The comparison shows a tiny offset of $+$0.014\,km\,s$^{-1}$ (APOGEE RVs minus {\it Gaia}), with a small scatter of 0.561\,km\,s$^{-1}$. The systematic trend of $\Delta \rm{RV}$ with color $G_{\rm{BP}}-G_{\rm{RP}}$, magnitude $G_{\rm{RVS}}$ and number of transits ($N_{\rm{obs}}$) is shown in Fig.~\ref{fig:Fig7}. Significant systematic trends for $\Delta \rm{RV}$ with color and magnitde are detected. For $G_{\rm{BP}}-G_{\rm{RP}}$, systematic deviations are clearly detected at $G_{\rm{BP}}-G_{\rm{RP}}<1$\,mag and $G_{\rm{BP}}-G_{\rm{RP}}>2$\,mag. For $G_{\rm{RVS}}$, systematic trend is significant found at $G_{\rm{RVS}}>11$\,mag. We first adopted a fourth-order polynomial to correct the trend along with $G_{\rm{RVS}}$. After corrections of $G_{\rm RVS}$ dependent systematics, the trend along with $G_{\rm{BP}}-G_{\rm{RP}}$ was further corrected by a fourth-order polynomial fit. The resulted coeficients are present in Table~\ref{table:Com-RV-STD}. It is worth noting that \citet{2022arXiv220605902K} has also identified the $G_{\rm RVS}$ dependent trend and provided a second-order polynomial to describe it for $G_{\rm{RVS}}>11$\,mag (as shown in Fig.~\ref{fig:Fig7}). The second-order polynomial can partially capture the systematic trend as we discovered, while it cannot correct the systematic trend at the faint end ($G_{\rm{RVS}}>13$\,mag).

Based on the common stars of our APOGEE RV standard star and {\it Gaia} DR3, we study the precision of {\it Gaia} RV measurements. As shown at the bottom panels of Fig.~\ref{fig:Fig7}, the precision of {\it Gaia} RV measurement generally decreases with $T_{\rm{eff}}$ and $G_{\rm{RVS}}$ and slightly increases with [Fe/H]. The RV precision at bright range ($\it{G}_{\rm{RVS}} < \rm{10}$\,mag) is several hundred m\,s$^{-1}$, and is few km\,s$^{-1}$ at the faint end ($\it{G}_{\rm{RVS}} > \rm{12}$\,mag). This result is consistent with the prediction of \citet{2022arXiv220605902K}.

\section{Summary}
\label{section:Sum}

We have constructed a catalog of 46,753 RV standard stars from the 657,000 near-infrared ($H$ band; 1.51--1.70\,$\mu$m) high-resolution ($R \sim$\,$22$\,$500$) spectra provided by APOGEE DR17. They are almost evenly distributed in the Northern and Southern Hemispheres, with 62\% red giants, and 38\% main-sequence dwarf stars. They were observed with a time baseline of at least 200 days (54\% longer than one year and 10\% longer than five years) and were observed more than 3 times. The median RV stability  was $215.25$\,m\,s$^{-1}$. Using the catalog of RV standard stars, we calibrated the RVZPs of four large-scale stellar spectroscopic surveys: the RAVE, LAMOST, GALAH and {\it Gaia}. By careful comparisons, we found the mean RVZPs are $+0.149$\,km\,s$^{-1}$, $+4.574$\,km\,s$^{-1}$ (for LRS), $-0.031$\,km\,s$^{-1}$ and $+0.014$\,km\,s$^{-1}$, for RAVE, LAMOST, GALAH and {\it Gaia}, respectively. In addition to an overall constant offset, RVZPs of part of these surveys show moderate dependences on stellar parameters (e.g., Teff, log\,$g$, [Fe/H], color or magnitude). We further provide corrections by simple polynomial fits with coefficients listed in Table~\ref{table:Com-RV-STD}. Our studies show that the small but clear RVZPs in these large-scale spectroscopic surveys can be well detected and properly corrected by our RV standard stars, which is believed to be useful for their further applications in various studies. The complete APOGEE RV standard star catalog in Table~\ref{table:RV-STD} is publicly available on the \url{https://nadc.china-vo.org/res/r101244/}.

\begin{acknowledgements}
This work is supported by National Key R \& D Program of China No. 2019YFA0405500, and National Natural Science Foundation of China grants 11903027, 11973001, 11833006, U1731108, 12090040, 12090044.

Funding for the Sloan Digital Sky Survey IV has been provided by the Alfred P. Sloan Foundation, the U.S. Department of Energy Office of Science, and the Participating Institutions. SDSS acknowledges support and resources from the Center for High-Performance Computing at the University of Utah. The SDSS web site is www.sdss.org.

\software{astropy \citep{2013A&A...558A..33A,2018AJ....156..123A}, TOPCAT \citep{2005ASPC..347...29T}}

\end{acknowledgements}


\bibliography{bibliography}{}
\bibliographystyle{aasjournal}

\end{document}